\documentclass[11pt,a4paper,twoside,groupcitations]{article}
\usepackage[T1]{fontenc}
\usepackage[ansinew]{inputenc}
\usepackage[english]{babel}
\usepackage{amsfonts}
\usepackage{amsmath}
\usepackage{bm}
\usepackage{array}
\usepackage{amsthm}
\usepackage{amssymb}
\usepackage{graphicx}
\usepackage{subfigure}
\usepackage{braket}
\usepackage{eucal}
\usepackage{verbatim}
\usepackage[table]{xcolor}
\usepackage{caption}
\usepackage{cite}
\usepackage{textcomp}
\usepackage{multicol}
\usepackage{tikz}
\usetikzlibrary{positioning,arrows}
\usetikzlibrary{decorations.pathmorphing}
\usetikzlibrary{decorations.markings}
\usetikzlibrary{calc,decorations.markings}
\usetikzlibrary{arrows,shapes}
\usetikzlibrary{matrix,arrows}
\usepackage{pgfplots}
\usepackage{xparse}
\definecolor{jade}{HTML}{00A86B}
\newcommand{\be}{\begin{eqnarray}}
\newcommand{\ee}{\end{eqnarray}}
\renewcommand{\d}{\mbox{${\rm d}$}} %d differenziale non corsivo in math mode

\newcommand{\gn}{G_{\rm N}}
\newcommand{\rh}{r_{\rm H}}

\title{\bf Quantum rotating black holes}
\author{Roberto~Casadio$^{ab}$\thanks{E-mail: casadio@bo.infn.it},
Andrea~Giusti$^{c}$\thanks{E-mail: agiusti@phys.ethz.ch}
$\ $and
Jorge~Ovalle$^{d}$\thanks{E-mail: jorge.ovalle@physics.slu.cz}
\\
\\
$^a${\em Dipartimento di Fisica e Astronomia, Universit\`a di Bologna}
\\
{\em via Irnerio~46, 40126 Bologna, Italy}
\\
\\
$^b${\em I.N.F.N., Sezione di Bologna, I.S.~FLAG}
\\
{\em viale B.~Pichat~6/2, 40127 Bologna, Italy}
\\
\\
$^c$ {\em Institute for Theoretical Physics, ETH Zurich}
\\
{\em Wolfgang-Pauli-Strasse 27, 8093 Zurich, Switzerland}
\\
\\
$^d$ {\em Research Centre of Theoretical Physics and Astrophysics}
\\
{\em 	Institute of Physics, Silesian University in Opava, CZ-746 01 Opava, Czech Republic}
}
\begin{document}
\maketitle
\begin{abstract}
We consider black holes generically sourced by quantum matter described by regular wavefunctions.
This allows for integrable effective energy densities and the removal of Cauchy horizons in 
spherically symmetric configurations.
Moreover, we identify the ultrarigid rotation of the Kerr spacetime as causing the existence of an
inner horizon in rotating systems, and describe general properties for quantum matter cores at the centre of
rotating black holes with integrable singularities and no Cauchy horizon.
\end{abstract}
\section{Introduction and motivation}
\setcounter{equation}{0}
\label{S:intro}
In the search for regular black holes one usually imposes regularity conditions inspired by 
classical physics, like finite (effective) energy density and scalar invariants (for recent reviews, see
Refs.~\cite{Maeda:2021jdc,Carballo-Rubio:2023mvr}).
These conditions allow one to remove the singularities that plague known black hole solutions~\cite{HE}
but usually bring back (or do not allow to remove) a seemingly undesirable inner Cauchy horizon. 
This fact is easily seen in the static spherically symmetric case, for which one can always introduce 
a Killing time $t$ and the areal radial coordinate $r$ in which the metric reads~\footnote{We use
units with $c=\gn=1$ and metric signature $(+---)$.}
\be
\d s^2
=
g_{tt}\,\d t^2
+
g_{rr}\,\d r^2 
-
r^2\left(\d \theta^2 +\sin^2 \theta \, \d \phi^2\right)
\ ,
\label{gsph}
\ee
where $g_{tt}=-e^{\varphi}\,g^{rr}$, with $\varphi=\varphi(r)$ a regular function and
\be
-g^{rr}
=
1
-
\frac{2\,m(r)}{r}
\ .
\ee
In the above, the Misner-Sharp-Hernandez mass function~\cite{Misner:1964je,Hernandez:1966zia} is given by
\be
m(r)
=
4\,\pi
\int_0^r
\epsilon(x)\,x^2\,\d x
\ ,
\label{msh}
\ee
where $\epsilon=\epsilon(r)$ is the proper energy density of the source.
We recall that the mass function approaches the ADM mass~\cite{ADM} $M$ in asymptotically flat space,
that is $m(r\to\infty)=M$.
For the Schwarzschild vacuum solution, $m=M$ and constant, and both the component $g_{tt}$ of the metric
and its Kretschmann scalar diverge for $r\to 0$. 
The latter result signals that tidal gravitational forces also diverge towards the centre.
\par
If we require that $\epsilon=\epsilon(r)$ is regular for $r\to 0$, we find $m\sim r^3$ and both the components of the
metric in the chosen frame and its Ricci and Kretschmann scalars remain finite in $r=0$~\cite{Maeda:2021jdc,Carballo-Rubio:2023mvr}. 
The central Schwarzschild singularity is removed but one necessarily finds a Cauchy horizon if there is 
an event horizon.
In fact, horizons are located at values of $r=\rh$ such that $g_{tt}(\rh)=g^{rr}(\rh)=0$.
Since $g_{tt}>0$ for $r>r_+=\rh$, it must be negative just inside the event horizon.
However, for $m\sim r^3$ one necessarily finds $g_{tt}(0)=e^{\varphi(0)}>0$, which implies that there must exist a second
zero $r=r_-$, with $0<r_-\le r_+$.~\footnote{The region of negative $g_{tt}$ shrinks to zero volume in the extremal case $r_-=r_+$.}
\par
In the quantum theory, one can allow for milder conditions to apply for the energy density of the source.
In particular, if one considers that
\be
\epsilon
\propto
|\Psi|^2
\ ,
\ee
where $\Psi=\Psi(r)$ is the wavefunction of the static matter source, the fundamental requirement is that $\Psi$ be integrable,
since a wavefunction must yield finite probability densities.
For any finite $r$, one then must have~\footnote{One might argue
that the correct volume measure must contain the determinant of the spatial metric, which implies an extra factor of $\sqrt{-g_{tt}}$
for $r<r_+$.
Such a factor would then depend on $\Psi$ itself according to Eq.~\eqref{msh}.
However, we notice that the condition~\eqref{Qcond} remains valid if $|g_{tt}|$ is finite everywhere inside $r_+$.}
\be
4\,\pi\,\int_0^r
|\Psi(x)|^2\,x^2\,\d x
<
\infty
\ . 
\label{Qcond}
\ee
This accommodates for the milder condition $\epsilon\sim r^{-2}$ and $m\sim r$, which still ensures that $m(0)=0$.
We shall recall in Section~\ref{S:QS} that this behaviour is tame enough to both replace the central singularity
with an integrable one and not allow for the emergence of a Cauchy inner horizon for electrically neutral
black holes~\cite{Casadio:2021eio}.
Moreover, the Cauchy horizon of the Reissner-Nordstr\"om black hole can also be eliminated by this
prescription~\cite{Casadio:2022ndh}.
Note that by {\em integrable singularity\/} we mean regions where the curvature invariants and the effective
energy-momentum tensor diverge, while their ``volume'' integrals remain finite~\cite{Lukash:2013ts}.
\par
When one considers rotating systems, however, the above condition on $\Psi$, hence on $m$, is not sufficient
to remove the inner horizon that appears in the Kerr black hole.
As we shall show in Section~\ref{S:spin}, this is due to the ultrarigid nature of the vacuum general relativistic solution,
which is characterised by a constant specific angular momentum $a=J/M$ (like its
generalisations~\cite{Newman:1965tw,gurses,Bambi:2013ufa}).
On the contrary, a radial dependent specific angular momentum is indeed more natural for extended bodies.
For example, an homogeneous sphere with mass function $m\sim r^3$ rotating with angular velocity $\omega$
has angular momentum $J\sim m\,r^2\,\omega\sim \omega\,r^5$ and $a\sim r^2$.
We will show that the simultaneous tempering of the ring singularity and removal of the inner horizon are accomplished by 
assuming that the quantum state $\Psi$ is such that $a\sim m\sim r$ for $r\to 0$.
\section{Spherical black holes} 
\label{S:QS}
\setcounter{equation}{0}
Consider a spherically symmetric and static spacetime described by the line element~\eqref{gsph} with $\varphi=0$,
that is
\be
\label{spherical-geo}
g_{tt}
=
-g^{rr}
=
1-\frac{2\,m(r)}{r}
\ ,
\ee
where the mass aspect $m=m(r)$ is not fixed a priori.
The corresponding energy-momentum tensor is obtained from 
\be
8\,\pi\,{T}_{\mu\nu} = G_{\mu\nu}
\ ,
\ee
with $G_{\mu\nu}$ denoting the Einstein tensor computed from the line element~\eqref{spherical-geo}.
By introducing the tetrad
\begin{eqnarray}
{e}^{\mu}_{t}
&\!\!=\!\!&
\left(\sqrt{g^{tt}},0,0,0\right)
\ ,
\qquad
{e}^{\mu}_{r}
=
\left(0,\sqrt{g_{tt}},0,0\right) \, ,
\nonumber
\\
\label{4sph}
\\
{e}^{\mu}_{\theta}
&\!\!=\!\!&
\left(0,0,r^{-1},0\right)
\ ,
\qquad
{e}^{\mu}_{\phi}
=
-\left(0,0,0,(r\,\sin\theta)^{-1}\right)
\ ,
\nonumber
\end{eqnarray}
the effective energy-momentum tensor can be written as
\be
\label{tmunu}
{T}^{\mu\nu}
=
{\epsilon}\, {e}^{\mu}_{t}\,{e}^{\nu}_{t}
+{p}_{r}\,{e}^{\mu}_{r}\,{e}^{\nu}_{r}
+{p}_{\theta}\,{e}^{\mu}_{\theta}\,{e}^{\nu}_{\theta}
+{p}_{\phi}\,{e}^{\mu}_{\phi}\,{e}^{\nu}_{\phi}
\ ,
\ee
with $\epsilon$ the energy density and $p_r$, $p_\theta$, and $p_\phi$ are pressure functions. 
One therefore finds 
\be
\epsilon
=
-p_r
=
\frac{m'(r)}{4\,\pi\,r^2}
\ ,
\label{Esph}
\ee
and
\be
p_\theta
=
p_\phi
=
-\frac{m''(r)}{8\,\pi\,r}
\ .
\label{Tsph}
\ee
Furthermore, from the metric functions in Eq.~\eqref{spherical-geo} one can easily compute the Ricci scalar
\be
R
=
-2\,\frac{r\,m''+2\, m'}{r^2}
\ ,
\label{RicciQS}
\ee
and the Kretschmann scalar
\be
R_{\alpha\beta\mu\nu}\,R^{\alpha\beta\mu\nu}
&\!\!=\!\!&
4\,\frac{r^4 \,(m'')^2+4 \left[2 \,r^2\, (m')^2-4\, r\, m\, m'+3\, m^2\right]+4\, r^2 \left(m-r\, m'\right) m''}{r^6}
\ .
\label{KrQS}
\ee
For $m\sim r^3$, it is easy to see that the Ricci scalar~\eqref{RicciQS} and the Kretschmann scalar~\eqref{KrQS},
as well as the effective energy density~\eqref{Esph} and pressures~\eqref{Tsph} remain finite for $r\to 0$.
The singularity is thus removed and tidal forces do not diverge by approaching what would be the singular
centre of the Schwarzschild black hole along radial geodesics.
\par
Under the weaker condition $m\sim r$, one can likewise check that all of the above quantities are still
(at least) integrable, and should therefore not be discarded in the quantum context.
Furthermore, it is easy to prove (see Appendix~\ref{App:Geodesics} for a detailed discussion)
that radial geodesics can be extended past $r=0$, namely the position of the would-be classical singularity.
These properties are clearly displayed by the models introduced in Refs.~\cite{Casadio:2021eio,Casadio:2022ndh}
and, {\em e.g.}, Refs.~\cite{Casadio:2020ueb,Brustein:2021lnr,Yokokura:2022kmq,cores}, but are not enjoyed by all regular black holes
in the literature~\cite{Zhou:2022yio}. In particular, the coherent quantum states in Refs.~\cite{Casadio:2021eio,Casadio:2022ndh} are meant to represent
(simplified) gravity states in any quantum theory of gravity, from which the effective classical geometry emerges
as a mean field for any matter source.
In this respect, it would be very interesting to investigate the transition from the more classical behaviour $m\sim r^3$
to $m\sim r$ which must happen in the interior of a collapsing body when it becomes a black hole.
\section{Rotating black holes} 
\label{S:spin}
\setcounter{equation}{0}
There exist several procedures to map a spherically symmetric metric into a rotating one,
like the Janis-Newman algorithm~\cite{Newman:1965tw}.
In the end, all of these procedures lead to a metric of the Kerr form with the constant ADM mass $M$
replaced by the mass function $m=m(r)$ of the spherically symmetric seed metric~\cite{gurses}. 
\subsection{The ultrarigid Kerr metric} 
Let us start by recalling that the Kerr metric reads
\be
ds^{2}
=
\left[1-\frac{2\,r\,M}{ {\rho}^2}\right]
dt^{2}
+
\frac{4\,  {a}\, r\,M\, \sin^{2}\theta}{ {\rho}^{2}}
\,dt\,d\phi
-
\frac{ {\rho}^{2}}{ {\Delta}}\,dr^{2}
-
 {\rho}^{2}\,d\theta^{2}
-
\frac{ {\Sigma}\, \sin^{2}\theta}{ {\rho}^{2}}\,d\phi^{2}
\ ,
\label{kerr}
\ee
where
\be
{\rho}^2
&\!\!=\!\!&
r^2+ {a}^{2}\cos^{2}\theta
\label{f0}
\\
 {\Delta}
&\!\!=\!\!&
r^2-2\,r\,M
+ {a}^{2}
\label{f2}
\\
{\Sigma}
&\!\!=\!\!&
\left(r^{2}+ {a}^{2}\right)^{2}
- {a}^{2}\,  {\Delta}\, \sin^{2} \theta
\label{f3}
\ee
and $a=J/M$, with $J$ is the angular momentum of the system.
\par
The above metric contains a ring singularity located at $\rho^2=0$ ({\em i.e.}, $r=0$ and $\theta = \pi/2$),
where the Kretschmann scalar $R_{\alpha\beta\mu\nu}\,R^{\alpha\beta\mu\nu}$ diverges (see Appendix~\ref{A:kerr}).
For $a^2< M^2$, the Kerr metric also has the horizons 
\be
r_\pm
=
M
\pm
\sqrt{M^2-a^2}
\ ,
\ee
corresponding to the two zeros of $\Delta=\Delta(r)$.
Note in particular that the existence of the inner (Cauchy) horizon follows from
the ``ultrarigid'' rotation described by a constant $a$.
In fact, one has
\be
\Delta(0)
=
a^2
>0
\ ,
\label{D>0}
\ee
which implies that $\Delta=\Delta(r)$ must change sign twice going inward, from positive to negative
across $\rh=r_+$ and then back to positive across $r_-$.
\par
\subsection{From spherical black holes to rotating black holes}
\label{S:RRBH}
We next consider the rotating metric obtained by replacing the constant $M$ in Eq.~\eqref{kerr} with a mass
function $m=m(r)$~\cite{gurses}.
In order to attenuate the ring singularity, one could impose $m\sim r^3$.
Like in the spherically symmetric case, the weaker condition $m\sim r$ is again sufficient to make
the relevant scalar quantities integrable around $r=0$.
However, it is not sufficient to remove the inner horizon, since Eq.~\eqref{D>0} still holds for $a$ constant.
\par
Let the mass function be analytic and related to an integrable energy density
in the neighbourhood of $r=0$, that is
\be
m
=
m_1\,r
+
M\,
\sum_{k=2}^\infty
m_k 
\left(\frac{r}{M}\right)^k
\ ,
\label{Sm}
\ee
where all the coefficients $m_1\ge 0$ and $m_k$ are dimensionless.
Further, we require a less than ultrarigid rotation by assuming that 
$a/\left(r/M\right)^\alpha$ is also analytic for fixed values of $\alpha\geq0$ in the neighbourhood of $r=0$.
In other words, we assume a specific angular momentum given by
\be
a
=
M\left(\frac{r}{M}\right)^\alpha
\sum_{k=0}^\infty
a_{k+1}
\left(\frac{r}{M}\right)^k
\ ,
\label{Sa}
\ee
in the neighbourhood of $r=0$, where the coefficients $a_k$ are also dimensionless.
Our metric will represent a black hole with one horizon at $r=\rh>0$ if $\Delta=\Delta(r)$
has {\em only one (strictly) positive real root} and $\Delta (0)\leq0$.
With the above expansions, on assuming $m_1\not=0$ and $a_1\not=0$, we have
\be
\Delta
\simeq
\left(\frac{r}{M}\right)^{2 \alpha }
\left(a_1\, M + a_2\, r\right)^2
-
\frac{2\, r^2}{M}
\left(m_1\, M + m_2\, r\right)
+
r^2
\ .
\ee
For $0\le \alpha < 1$, we would have $\Delta \simeq (r/M)^{2\alpha}\,a_1^2\,M^2$,
which is always positive near $r=0$ since $a_1 \neq 0$.
If we assume $\alpha = 1$, then $\Delta \simeq (1 + a_1^2 - 2\, m_1)\,r^2$, which
is negative inside the horizon provided 
\be
2\, m_1 > 1 + a_1^2
\ .
\label{c1}
\ee
Lastly, $\Delta\simeq (1 - 2\, m_1)\, r^2$, for $\alpha>1$, which is consistent
with the condition $\Delta < 0$ inside the horizon if $2\,m_1>1$.
Note that for $\alpha\ge 1$, we always have $\Delta(0)=0$, so that the location of
the would-be inner Cauchy horizon is in fact pushed to the centre of the system.
\par
\begin{figure}
\centering
\includegraphics[scale=0.5]{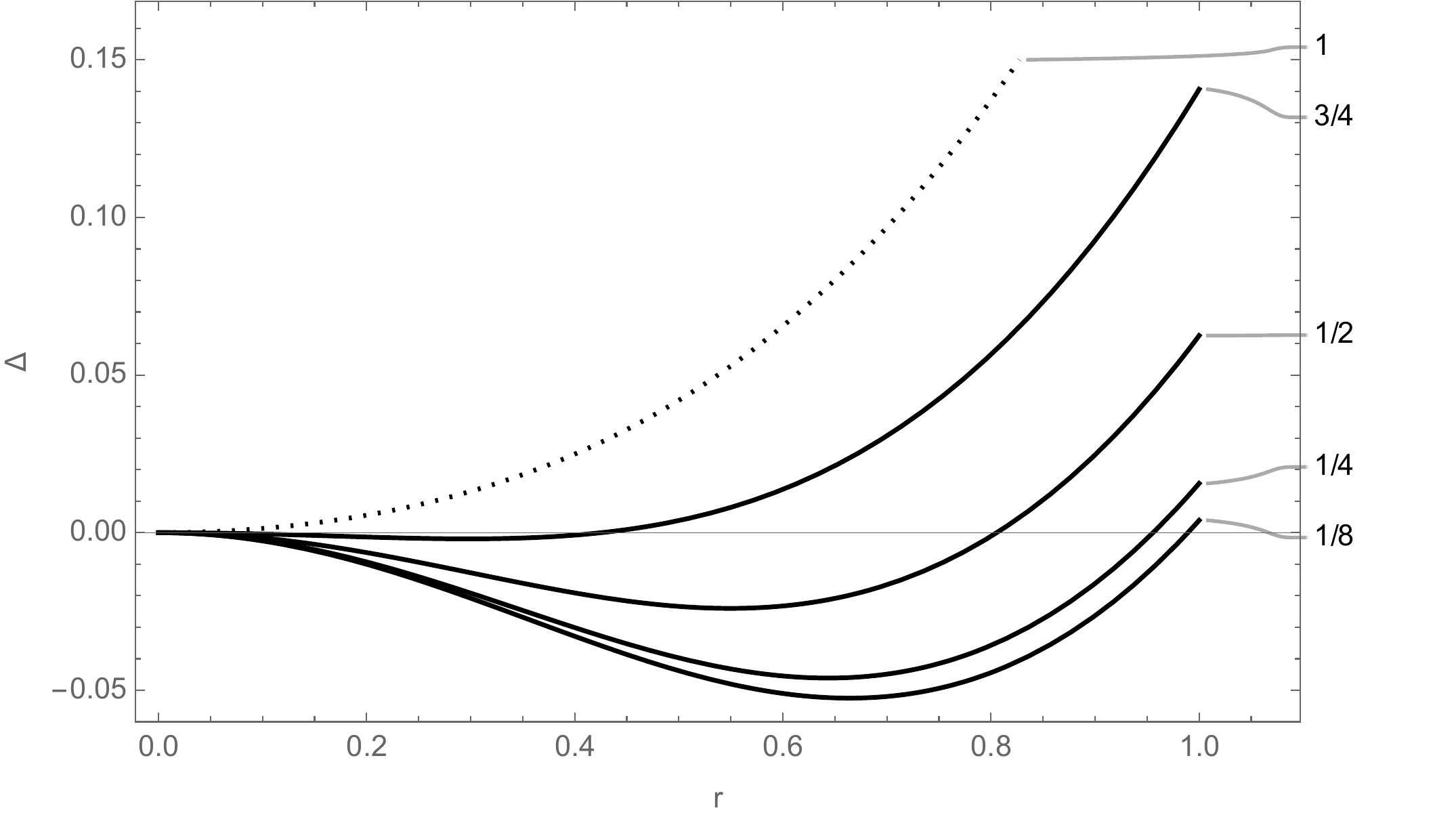}
\caption{$\Delta=\Delta(r)$ in Eq.~\eqref{example} for $M=1$ for different values of $A$ satisfying the condition~\eqref{c1} (solid lines)
and a case violating Eq.~\eqref{c1} (dotted line).
All solid lines display one horizon where $\Delta(\rh>0)=0$.
\label{Fig1}}
\end{figure}

To clarify the above result, let us consider a simple example given by
\be
m
=
\frac{2\, M}{\pi} \, \arctan\!\left( \frac{r}{ M} \right)
\quad
\mbox{and}
\quad
a
=
\frac{2\,A}{\pi} \, \arctan\!\left( \frac{r}{M} \right)
\ ,
\ee
which smoothly interpolate between the required behaviour near $r=0$
and the asymptotic ADM mass $M$ and specific angular momentum $A$.
In particular,
\be
m
=
\frac{2}{\pi} \, r +\mathcal{O} (r^2) \quad \mbox{for} \ r\to 0^+
\ , 
\quad
\lim_{r \to \infty} m(r) = M
\ ,
\ee
which implies $m_1=2/\pi$, and
\be
a
=
\frac{2\,A}{\pi\,M}\,r
+\mathcal{O}(r^2)
\quad
\mbox{for}\ 
r\to 0^+
\ , 
\quad
\lim_{r \to \infty} a(r) = A
\ ,
\ee
which yields $a_1=2\,A/\pi\,M$.
Furthermore, one finds that
\be \label{example}
\Delta
=
r^2 +\frac{4 \,A^2}{\pi ^2}\, \arctan^2\!\left( \frac{r}{ M} \right)
-\frac{4\, M\, r}{\pi}\, \arctan\!\left( \frac{r}{ M} \right)
\ , 
\ee
which admits at most one horizon at $\rh>0$ for values of $A$ satisfying the condition~\eqref{c1},
that is $0<A<\sqrt{(1-\pi/4)\,\pi}\,M$, as is shown in Fig.~\ref{Fig1}.
\par
The effective energy-momentum tensor for a rotating metric with general $m=m(r)$ and $a=a(r)$ is given in Appendix~\ref{A:axis}.
In particular, the behaviour near the centre for mass and specific angular momentum of the forms in Eqs.~\eqref{Sm} and \eqref{Sa}
with $\alpha=1$ is analysed in Appendix~\ref{A:center}.
\section{Concluding remarks} 
\label{S:conc}
\setcounter{equation}{0}
In Refs.~\cite{Casadio:2021eio} and~\cite{Casadio:2022ndh} coherent states were constructed to reproduce
the classical Schwarzschild and Reissner-Nordstr\"om metrics.
The condition of normalisability of such quantum states implies the necessary departure from the respective exact 
classical geometries, an effect that one can understand with the existence of an extended (quantum) matter
core~\cite{Casadio:2021cbv,Casadio:2022pla,Casadio:2020ueb,Yokokura:2022kmq,cores}.
The similar quantum condition~\eqref{Qcond} has been analysed here in more generality to show that the classical
singularity can be generically replaced by an integrable singular structure in the interior of a spherically symmetric
black hole if $m\sim r$ near $r=0$.
\par
For a spherically symmetric system, the above condition on the mass function can be implemented in order to also avoid
the presence of inner horizons (see Ref.~\cite{Casadio:2022ndh} for the details).
However, the inner horizon in rotating black holes can be circumvented only if one further assumes that 
the specific angular momentum $a$ is not constant throughout space but vanishes sufficiently fast towards
the centre.
This additional condition is in fact natural if one considers that the geometry is again sourced by a quantum
core~\cite{Casadio:2022epj}, for even classical extended bodies do not rotate as rigidly as the vacuum Kerr spacetime.
We showed that the conditions $m\sim a\sim r$ are sufficient to ensure an integrable interior without Cauchy horizon.
One could therefore use the quantum metrics from Refs.~\cite{Casadio:2021eio,Casadio:2022ndh}
as seeds for generating the rotating integrable versions of the Kerr and Kerr-Newman geometries by means
of the algorithms described in Refs.~\cite{gurses,Contreras:2021yxe} with the additional generalisation $a\sim m$
(or faster decay towards the centre). 
\par
The absence of Cauchy horizons cures the issue of mass inflation that plagues most regular black hole
geometries present in the literature~\cite{Carballo-Rubio:2023mvr}.
This suggests that the quantum picture of black holes developed in this work could indeed represent
(meta)stable final configurations for the gravitational collapse of compact astrophysical objects.
Since our analysis is limited to stationary configurations, the effects due to matter accretion,
or the merging of two black holes, and the Hawking evaporation must be left for future developments.
In particular, we notice that the global size of the core is not relevant for the issues discussed here,
as long as $m\sim a\sim r$ in a sufficiently large neighbourhood of the centre.
However, we expect that departures from the outer classical geometry (usually termed ``quantum hair'')
are highly sensitive to the relative size of the core radius with respect to the horizon radius, as shown explicitly in
Refs.~\cite{Casadio:2021eio,Casadio:2022ndh}, and that all of black hole phenomenology is therefore
related to this aspect of the interior.
\section*{Acknowledgments}
R.C.~is partially supported by the INFN grant FLAG.
A.G.~is supported by the European Union's Horizon 2020 research and innovation programme
under the Marie Sklodowska-Curie Actions (grant agreement No.~895648-CosmoDEC).
The work of R.C.~and A.G.~has also been carried out in the framework of activities of the
National Group of Mathematical Physics (GNFM, INdAM).
J.O.~is partially supported by ANID FONDECYT grant No.~1210041.
%
%\section*{Data availability statement}
%
%Data sharing not applicable to this article as no datasets were generated or analysed during the current study.
%
%
\appendix
\section{Geodesics in spherical spacetime}
\label{App:Geodesics}
Let us consider the Lagrangian for the motion of a point-like particle in the metric~\eqref{gsph} with $\varphi=0$,
\be
\mathcal{L}
=
\left(1-\frac{2\,m}{r}\right)\dot t^2
-\left(1-\frac{2\,m}{r}\right)^{-1}\dot r^2
-r^2 \, \dot \theta^2
-r^2 \, \sin^2 \theta\, \dot \phi^2
\ ,
\ee
where dots denote derivatives with respect to the an affine parameter $\lambda$.
This Lagrangian is constant along geodesics and takes the value $\mathcal{L}=1$ for timelike trajectories
and $\mathcal{L}=0$ for the null case. 
Since $\mathcal{L}$ does not depend explicitly on $t$ and $\phi$, the corresponding conjugate momenta
are conserved and read
\be
E
=
g_{tt} \, \dot{t}
=
\left[1 - \frac{2 \, m(r)}{r}\right] \dot{t} 
\quad
\mbox{and}
\quad 
L= r^2 \sin^2\theta \, \dot \phi
\ . 
\ee
Furthermore, taking advantage of the spherical symmetry one can fix $\theta = \pi/2$ for all $\lambda$,
without loss of generality.
By plugging in the conserved quantities $E$ and $L$, one finds the radial equation
\be
\label{dotlambda}
\frac{\dot r ^2}{2} + V_{\rm eff}
=
\frac{E^2}{2}
%=:
%\bar{E}
\ ,
\ee
with
\be
V_{\rm eff}
=
\frac{1}{2} \left[ 1 - \frac{2 \, m(r)}{r} \right]\left(\mathcal{L} + \frac{L^2}{r^2}\right)
\ .
\ee
\par
For radial geodesics with $L=0$, and assuming $m = m_1\,r + \mathcal{O}(r^2)$ as $r \to 0^+$, we find
\be
V_{\rm eff}
=
-\left(m_1-\frac{1}{2}\right)\mathcal{L}
+
\mathcal{O}(r) 
\quad
\mbox{for}\ r \to 0^+
\ ,
\ee
which guarantees the extendibility of radial geodesics past $r=0$.
Furthermore, $g_{tt}$ must change sign only once (from positive to negative inward), say at $r=\rh$,
in order to guarantee the existence of one horizon.
This implies that the component $g_{tt}$ cannot be positive at $r=0$ and the condition $m_1\ge 1/2$.
\par
We should remark that a process of gravitational collapse leading to a spherically symmetric configuration
must satisfy the same degree of symmetry throughout its evolution.
In other words, in order to potentially form the geometry~\eqref{gsph}, a self-gravitating system
must undergo a spherical collapse.
As a result, the relevant causal structure for the system is determined by radial null geodesics.
Geodesics with non-vanishing angular momentum are of limited interest since
the presence of matter moving along such trajectories would require to abandon the spherical
symmetry in order to provide a fully consistent description, particularly inside the core sourcing the geometry.
\section{Effective energy-momentum tensor} 
\label{A:axis}
\setcounter{equation}{0}
For a metric of the form~\eqref{kerr} with mass function $M=m(r)$ and $a=a(r)$,
it is convenient to introduce the tetrads~\cite{Misner:1973prb}
\begin{eqnarray}
{e}^{\mu}_{t}
&\!\!=\!\!&
\frac{\left(r^{2}+{a}^{2},0,0,{a}\right)}{\sqrt{\rho^{2}\Delta}}
\ ,
\qquad
{e}^{\mu}_{r}
=
\frac{\sqrt{\Delta}\left(0,1,0,0\right)}{\sqrt{\rho^{2}}}
\nonumber
\\
\label{4}
\\
{e}^{\mu}_{\theta}
&\!\!=\!\!&
\frac{\left(0,0,1,0\right)}{\sqrt{\rho^{2}}}
\ ,
\qquad
{e}^{\mu}_{\phi}
=
-\frac{\left({a}\sin^{2}\theta,0,0,1\right)}{\sqrt{\rho^{2}}\sin\theta}
\ ,
\nonumber
\end{eqnarray}
so that the effective source can be written in the form~\eqref{tmunu}, 
where the energy density is given by
\be
\epsilon
&\!\!=\!\!&
\frac{r^2\,m'}{4\,\pi\,\rho^4}
-
\frac{r+m-r\,m'
-
\left(3\,r+r\,m'-9\,m\right)
\cos^2\theta}
{8\,\pi\,\rho^6}\,
r^2\,a\,a'
\nonumber
\\
&&
+
\frac{3\,r+m+r\,m'
-
\left(r-m-r\,m'\right)
\cos^2\theta}
{8\,\pi\,\rho^6}\,
a^3\,a'\,\cos^2\theta
\nonumber
\\
&&
-\frac{
9\,r^2+14\,r\,m
+
8\left(
2\,r^2+a^2
\right)
\cos^2\theta
-
\left(
r^2-2\,m\,r
\right)
\cos 4\theta}
{64\,\pi\,\rho^6}
\,a^2\,(a')^2
\nonumber
\\
&&
-
\frac{r^2
-3\,m^2
+
\left(r^2+3\,m^2-4\,r\,m\right)
\cos^2\theta}
{8\,\pi\,\rho^6}
\,r^2\,(a')^2
\nonumber
\\
&&
-
\frac{\Delta\,(1+\cos^2\theta)}{8\,\pi\,\rho^4} \,a\,a''
\ ,
\ee
and the radial pressure reads 
\be
{p}_r
&\!\!=\!\!&
-\frac{r^2\,m'}{4\,\pi\,\rho^4}
+
\frac{r^3+r^2\,m-r^3\,m'
+
\left(
r-m-r\,m'
\right)
a^2\,\cos^2\theta}
{8\,\pi\,\rho^6}
\left(1+\cos^2\theta\right)
a\,a'
\nonumber
\\
&&
+
\frac{
r^2\,m^2
+
\left[
\left(r^2
+2\,r\,m
\right)
a^2
-r^2\,m^2
\right]
\cos^2\theta
+
a^4\,\cos^4\theta}
{8\,\pi\,\rho^6}\,
(a')^2
\ .
\ee
The two tensions are given by
\be
{p}_\theta
&\!\!=\!\!&
-\frac{a^2\,m'\,\cos^2\theta}{4\,\pi\,\rho^4}
-
\frac{r^3+r^2\,m-r^3\,m'
+
\left(
r-m-r\,m'
\right)
a^2\,\cos^2\theta}
{8\,\pi\,\rho^6}
(1+\cos^2\theta)
\,a\,a'
\nonumber
\\
&&
+
\frac{r^4-r^2\,m^2
+\left[
\left(
r^2-2\,r\,m
\right)
a^2
-r^2\,m^2
\right]
\cos^2\theta}
{8\,\pi\,\rho^6}\,
(a')^2
\nonumber
\\
&&
-\frac{r\,m''}{8\,\pi\,\rho^2}
+
\frac{a\,a''}{8\,\pi\,\rho^2}
\ee
and 
\be
{p}_\phi
&\!\!=\!\!&
-\frac{a^2\,m'\,\cos^2\theta}{4\,\pi\,\rho^4}
+
\frac{
r-3\,m+3\,r\,m'
-
\left(3\,r-m+r\,m'\right)\cos^2\theta}
{8\,\pi\,\rho^6}
r^2\,a\,a'
\nonumber
\\
&&
-
\frac{
\left(r-m-r\,m'\right)
\left(3-\cos^2\theta\right)}
{8\,\pi\,\rho^6}
a^3\,a'\,\cos^2\theta
\nonumber
\\
&&
+
\frac{4\,r^4\,\cos^2\theta+4\,r^2\,m^2\,\sin^2\theta-a^4\,\sin^2 2\theta}
{32\,\pi\,\rho^6}
(a')^2
\nonumber
\\
&&
+
\frac{(4\,\cos 2\theta+\sin^2 2\theta)\,r
-(9-\cos 4\theta)\,m}
{32\,\pi\,\rho^6}
\,r\,a^2\,(a')^2
\nonumber
\\
&&
-\frac{r\,m''}{8\,\pi\,\rho^2}
+
\frac{2\,r\,m-a^2
+\left(
r^2-2\,r\,m-2\,a^2
\right)
\cos^2\theta}
{8\,\pi\,\rho^4}\,
a\,a''
\ ,
\ee
and we further notice that $p_\theta=p_\phi$ on the axis (that is, for $\theta=0$ or $\pi$).
\par
Other quantities of interest are the Ricci scalar
\be
R
&\!\!=\!\!&
2\,\frac{2\, a\,a''+(a')^2-r \,m''-2 \,m'}{\rho^6}\,a^4\,\cos ^4\theta
-
2\,
\frac{r\, m''+2\, m'}{\rho^6}
\,r^4
\nonumber
\\
&&
+
2\,
\frac{\left[3\,r+(r-2\,m)\,\sin^2\theta\right](a')^2\cos^2\theta
-\left(r\, m''+2 \,m'\right)\sin^2\theta}
{\rho^6}
\,r\,a^2
\nonumber
\\
&&
+
\frac{r \,a'' \left[r \left(3+\cos 2\theta\right) -4\,m\, \cos ^2\theta\right]
-2 \,a' \left[2\,r -r\,(3+m')\,\sin^2\theta-\left(4-5\,\sin^2\theta\right) m\right]}
{\rho^6}
\,r^2\,a
\nonumber
\\
&&
+
\frac{r \,a'' \left[r \left(7+\cos2 \theta\right)-4\,m\, \cos ^2\theta\right]
-2\,a' \left[2\,r +r\,(1-m')\,\sin^2\theta-m\,\sin^2\theta\right]}
{\rho^6}
\,a^3\cos ^2\theta
\nonumber
\\
&&
+
2\,\frac{2\,r-(r-m)\,\sin^2\theta}
{\rho^6}
\,r^2\,(a')^2
\ ,
\ee
and the Kretschmann scalar $R_{\alpha\beta\mu\nu}\,R^{\alpha\beta\mu\nu}$, whose expression is really too cumbersome to display.
\par
For consistency, we will show below that the expressions for constant $a$ and $m$ are correctly recovered.
\subsection{Kerr-like case} 
\label{A:kerr}
The above expressions reduce to those given in Ref.~\cite{Contreras:2021yxe}
for $a(r)=a$ constant, that is
\be
\epsilon
=
-p_r
=
\frac{r^2\,m'}{4\,\pi\,\rho^4}
\ee
and
\be
p_\theta
=
p_\phi
=
-\frac{r\,m''}{8\,\pi\,\rho^2}
-\frac{a^2\,\cos^2\theta\,m'}{4\,\pi\,\rho^4}
\ .
\ee
Moreover, the Ricci scalar reads
\be
R
=
-2\,\frac{r\,m''+2\, m'}{\rho^2}
\label{RicciQK}
\ee
and 
\be
R_{\alpha\beta\mu\nu}\,R^{\alpha\beta\mu\nu}
&\!\!=\!\!&
4\, \frac{r^2 (m'')^2}
{\rho^4}
+
16\,
\frac{m' \left(2\,r^4-5\, a^2\, r^2 \cos ^2\theta+a^4 \cos ^4\theta\right)
-r \,m'' \left(r^4-a^4 \cos ^4\theta\right)}
{\rho^{8}}
\,m'
\nonumber
\\
&&
+
16\,
\frac{r^2-3 \,a^2 \cos ^2\theta}
{\rho^{8}}
\,r^2\,m\,m''
-
64\,
\frac{r^4-8 \,a^2\, r^2 \,\cos ^2\theta+3\, a^4\, \cos ^4\theta}
{\rho^{10}}
\,r\,m\,m'
\nonumber
\\
&&
+
48\,m^2\,
\frac{r^6-15\, a^2 \,r^4 \cos ^2\theta+15 \,a^4 \,r^2 \cos ^4\theta-a^6 \,\cos ^6\theta}
{\rho^{12}}
\ .
\label{KrQK}
\ee
For the standard Kerr solution, one of course has the Ricci scalar $R=0$ and
\be
R_{\alpha\beta\mu\nu}\,R^{\alpha\beta\mu\nu}
=
48\,M^2\,
\frac{r^6-15\,a^2\,r^4\,\cos^2\theta+15\,a^4\,r^2\,\cos^4\theta-a^6\,\cos^6\theta}
{\rho^{12}}
\ ,
\ee
which diverges for $r=0$ at the equator $\theta=\pi/2$.
\subsection{Central quantities} 
\label{A:center}
We can finally study the leading order behaviour of the energy-momentum tensor near $r=0$
on assuming the expansions~\eqref{Sm} and \eqref{Sa} with $\alpha=1$.
Since the general expressions are very cumbersome, we shall just consider the effective
energy density and pressure terms on the equator ($\theta=\pi/2$) and on
the axis of symmetry ($\theta=0$).
\par
For $\theta=\pi/2$, the leading terms in $r/M$ for the energy density are given by
\be
\epsilon
&\!\!\simeq\!\!&
\frac{2\,m_1
-
\left(1+2\,m_1\right)a_1^4
-
\left(2-3\,m_1^2\right)
a_1^2}
{8\,\pi\,r^2}
\nonumber
\\
&&
+
\frac{m_2
\left(4+a_1^2-6\,m_1\,a_1^2-2\,a_1^4\right)
-\left(9-4\,m_1-12\,m_1^2+12\,m_1\,a_1^2+8\,a_1^2\right)a_1\,a_2}{8\,\pi\,M\,r}
\ .
\ee
The radial pressure reads
\be
p_r
\simeq
-\frac{2\,m_1
-
\left(1+m_1^2\right)a_1^2}
{8\,\pi\,r^2}
-
\frac{m_2\left(4-2\,m_1\,a_1^2+a_1^2\right)
-\left(3+4\,m_1^2\right)a_1\,a_2}
{8\,\pi\,M\,r}
\ ,
\ee
the tensions
\be
p_\theta
\simeq
-\frac{a_1^2\,m_1^2}
{8\,\pi\,r^2}
-
\frac{2\,m_2
-\left(1-2\,m_1\right)
m_1\,a_1^2
-\left(3-4\,m_1^2\right)a_1\,a_2}{8\,\pi\,M\,r}
\ee
and
\be
p_\phi
&\!\!\simeq\!\!&
\frac{a_1^2\left(1+m_1^2-2\,m_1\,a_1^2-a_1^2\right)}
{8\,\pi\,r^2}
\nonumber
\\
&&
-
\frac{m_2
\left(2-3\,a_1^2-2\,m_1\,a_1^2+2\,a_1^4\right)
-\left(3+4\,m_1+4\,m_1^2-12\,m_1\,a_1^2-8\,a_1^2\right)a_1\,a_2}{8\,\pi\,M\,r}
\ .
\ee
\par
For $\theta=0$, we find 
\be
\epsilon
&\!\!\simeq\!\!&
\frac{2\,m_1
\left(1-a_1^2+a_1^4\right)
-a_1^4-a_1^6}
{8\,\pi\,(1+a_1^2)^3\,r^2}
\\
&&
+
\frac{m_2\,(1+a_1^2)\,
\left(2+a_1^2+2\,a_1^4\right)
-\left[3+2\,m_1\left(2-8\,a_1^2+a_1^4\right)+11\,a_1^2\left(1+a_1^2\right)+3\,a_1^6\right]a_1\,a_2}
{4\,\pi\,(1+a_1^2)^4\,M\,r}
\ ,
\nonumber
\ee
\be
p_r
&\!\!\simeq\!\!&
-\frac{2\,m_1
-
a_1^2
\left(1-a_1^2\right)
\left(2-2\,m_1+a_1^2\right)}
{8\,\pi\,(1+a_1^2)^3\,r^2}
\nonumber
\\
&&
-
\frac{m_2\left(2+5\,a_1^2+5\,a_1^4+2\,a_1^6\right)
-\left[3+2\,m_1\left(2+a_1^4\right)+5\,a_1^2+3\,a_1^4+a_1^6\right]a_1\,a_2}
{4\,\pi\,(1+a_1^2)^4\,M\,r}
\ee
and
\be
p_\theta
=
p_\phi
&\!\!\simeq\!\!&
-\frac{a_1^2
\left(1+2\,m_1+a_1^2\right)}
{8\,\pi\,(1+a_1^2)^3\,r^2}
\nonumber
\\
&&
-
\frac{m_2\left[1+a_1^2\left(2+a_1^2\right)^2\right]
+\left[2\,m_1\left(1-2\,a_1^2\right)-3\,a_1^2-4\,a_1^4-a_1^6\right]a_1\,a_2}
{4\,\pi\,(1+a_1^2)^4\,M\,r}
\ .
\ee
\end{document}